# IR- and visible light single photon detection in superconducting $MgB_2$ nanowires.

*Sergey Cherednichenko*, Narendra Acharya, Evgenii Novoselov, and Vladimir Drakinskiy*

Terahertz and Millimetre Wave Laboratory, Department of Microtechnology and Nanoscience, Chalmers University of Technology, SE-412 96, Gothenburg, Sweden

* *Corresponding author serguei@chalmers.se*

**Abstract**. State-of-the-art Superconducting Nanowire Single Photon Detectors based on low-$T_c$ materials reach 100% quantum efficiency. However, the response time is limited to >1-10 ns. Recently, it has been shown that due to a much lower kinetic inductance, a 100 ps response rate can be achieved in 120 μm-long $MgB_2$ nanowires. In this work, we demonstrate experimentally that such $MgB_2$ nanowires function as single- photon detectors for both visible (λ= 630 nm) and infrared (λ= 1550 nm) photons when biased close to the critical current, with a dark count rate of <10 cps. $MgB_2$ photodetectors over- perform NbN SNSPDs in speed by at least an order of magnitude for similar nanowire lengths. Such photodetectors offer a platform for single-photon detectors with a long thought- after combination of a large detector area and a response rate of up to 10 GHz with a single readout line.

## Main Text

Ability to detect single photons opens many interesting applications from quantum physics to space communication. [1,2,3] Recently, Superconducting *Nanowire* Single Photon Detectors (SNSPD or SSPD) [4,5] have shown a combination of both high efficiency and count rates up to 100 MHz in single photon detection from the visible to the mid-IR ranges. [6,7,8] Absorption of a photon in a superconducting nanowire ($w$~100 nm wide) triggers a normal-state domain growth, which can (under favorable conditions) stretch across the entire nanowire width.[9] This causes a voltage drop of a magnitude sufficiently large to be registered with standard microwave amplifiers and pulse counters. The aforementioned favorable conditions include the presence of a dc bias current, $I_0$, flowing through the nanowire close to the superconducting critical current, $I_c$ (sometimes called a switch current). Ratios of $I_0/I_c$ close to 1 are required in order to achieve single photon sensitivity and increase the quantum efficiency. [4,6] After the normal domain formation, the bias current is partly rerouted from the nanowire into the readout load, $R_L$=50 Ω. The nanowire detector resets only after the normal domain has collapsed and the bias current has returned back into the nanowire. Whereas the former process is governed by electron-phonon cooling, the later process is mainly limited by the large kinetic inductance in the nanowire,[10] which is normally long in order to cover a large detection area. For the most common SNSPD materials, such as NbN and NbTiN, a kinetic inductance of $L_{k0}$~90 pH/□ has been reported, which results in a reset time of $\tau = L_{k0} \times l / (w \times R_L)$ ~3-5 ns for ~$l$ = 100-150 μm long devices. [6,8] On the other hand, we recently demonstrated that in nanowires made from 5nm-thick $MgB_2$ films a much lower kinetic inductance, $L_{k0}$~1.5 pH/□ , can be achieved.[11] Combined with a 12 ps electron relaxation time [12], this has resulted in a reset time of ~100ps for $MgB_2$ devices as long as 120 μm. In this paper, we studied photon detection in $MgB_2$ nanowires in both the visible and the laser- communication (1550 nm) ranges. We demonstrated that such detectors generate responses from 1-, 2, or 3- photon absorptions depending on the bias current. In a comparative experiment, $MgB_2$ nanowire photodetectors demonstrate a response rate of at least an order of magnitude shorter than that in NbN SNSPDs.

For this study, $MgB_2$ thin films were grown on 6H-SiC substrates using Hybrid Physical Chemical Vapor Deposition (HPCVD). Both the film growth process and the nanowire fabrication routines have been described in earlier publications. [11],[13],[14] $MgB_2$ films were ~5 nm-thick, as has been shown by Transmission Electron Microscopy,[11] with a critical temperature $T_c$ of ~32-34 K (Fig.1a). All photon detection experiments were conducted in a cryogen-free probe station (base temperature 5 K) with an optical view port and a microwave ground- signal- ground (GSG) probe on a XYZ-moving stage, which provided both the dc biasing and the broadband readout. Schematics of the experimental set-up are shown in Fig.2a. A 630nm-line cw laser and a 1550nm-line 80 fs pulsed laser (100MHz repetition rate) flood illuminated the samples through a set of neutral density filters (visible or IR), a quartz sealing window of the probe station and an IR filter (transparent for wavelengths <2μm) at the 60 K heat shield. At 5K, the critical currents were ~97μA and ~399μA for a 35nm-wide and 100nm-wide $MgB_2$ nanowires (Fig.1b). For a direct comparison of the response rate of $MgB_2$ and NbN nanowire detectors, we fabricated a set of 100nm-wide NbN nanowire detectors, 115 μm and 460 μm long. NbN films were grown on C-cut sapphire substrates using dc magnetron sputtering at a substrate temperature of 800 °C. Other details could be found in the supplementary material. The NbN film was ~5nm thick, as estimated from the deposition rate. The critical temperature was ~10 K (Fig.1a) and the room temperature sheet resistance was ~300 Ω/□, which corresponded fairly well to previously published NbN SNSPDs. Forms of current- voltage ( $I(V)$) curves for NbN samples were quite similar to those for MgB2 samples, with a distinct switching behavior at $I_c$ , without latching. However, the critical current density in MgB2 samples showed a factor of 10-12 times higher than that shown in NbN samples (Fig.1b). All samples were integrated into coplanar waveguide contacts suitable for the 100µm pitch GSG-probe of the cryo-station. At room temperature, both microwave readout and dc biasing were arranged (Fig.2a) though a bias-T (Picosecond 5547, 12kHz-15GHz) followed by a low noise amplifier (MITEQ 3D-0010025, 0.5-5 GHz) and either a real time oscilloscope (Keysight Infiniium 54854 DSO, 4GHz, 20 GSa/s) or a pulse counter (HP 53131A 225). Nanowires were biased with a Yokogawa 7651 programmable dc source.

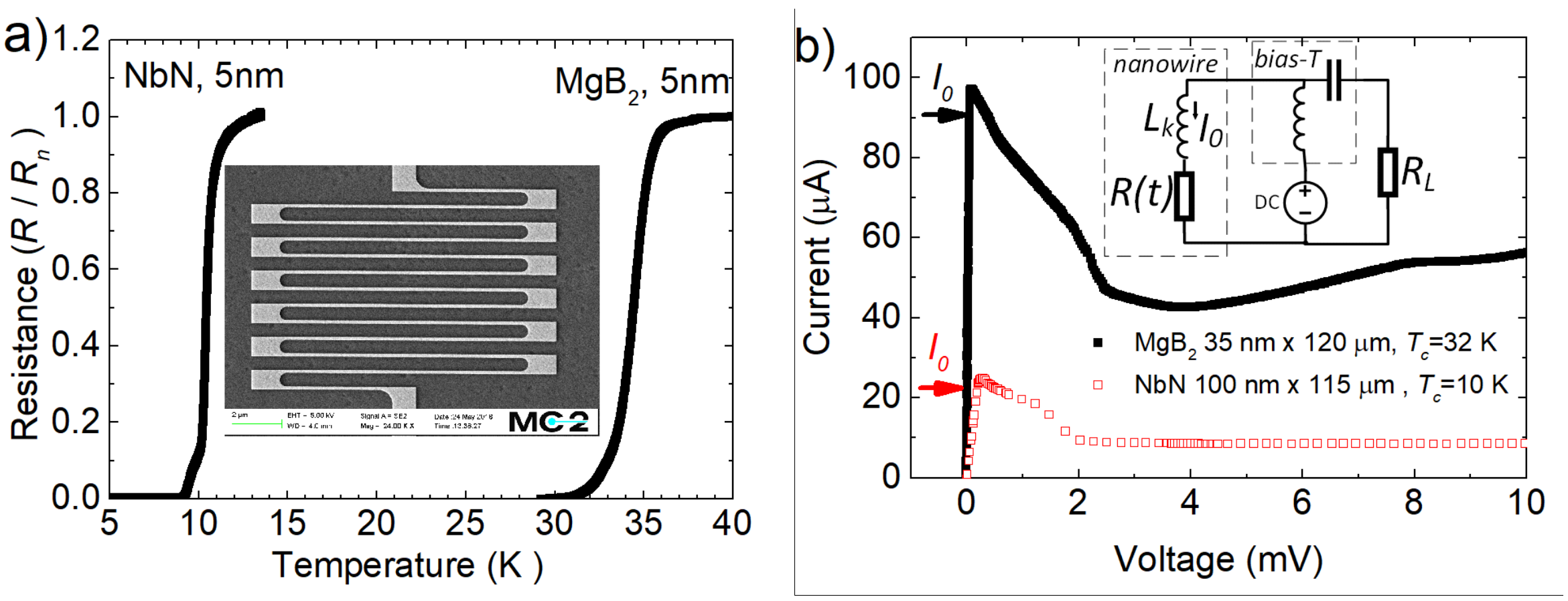

Fig. 1 (a) Superconducting transitions (normalized resistance) in a MgB2 and a NbN nanowire. The inset shows a SEM image of a 120 µm-long MgB2 nanowire. (b) $I(V)$-curves (at 5K) for a MgB2 and a NbN nanowire, utilized for the response rate experiments (see Fig.2b). Bias points for each photodetector are indicated with arrows. The inset shows simplified schematics of a photo detector (a variable resistance and an inductor, representing kinetic inductance in the nanowire) with a biasing network and a readout load.

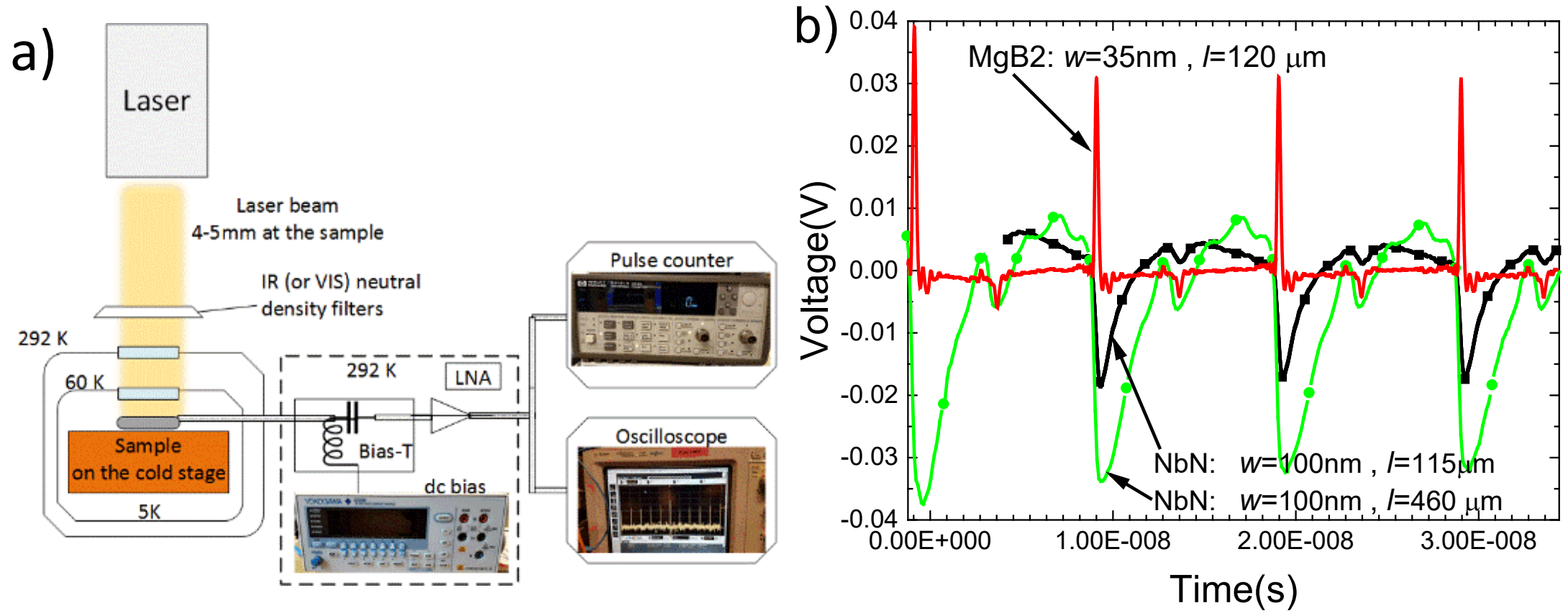

Fig. 2 (a) The optical characterization set-up. The fs-laser (or the cw laser) flood illuminates the sample on the 5 K cooling stage through the window in the cryo probe-station. (b) Response to the IR pulsed laser (1550nm, 100MHz repetition rate) of two NbN samples (100nm×115µm (squares) and 100nm×460µm (circles)) and one $MgB_2$ sample (35nm×120µm (thin red)). The same readout chain was used for all detectors (0.05-5GHz LNA + 4GHz oscilloscope). The voltage polarity for the $MgB_2$ nanowire response was inverted for easier reading of the figure.

The voltage response of two types of NbN devices (100 nm × 115 µm and 100 nm × 460 µm) and one $MgB_2$ device (35 nm × 120 µm) to excitations by the 80fs pulsed IR laser are shown in Fig.2b. All three devices were tested with exactly the same readout circuit, in a single cool-down of the cryogenic probe-station. The voltage decay in NbN nanowires (3.7ns for the 460 µm-long sample and 1.3ns for the 115µm-long sample) corresponds to the measured kinetic inductances of 200nH and 54nH, respectively (see supplementary material). The voltage decay time for the $MgB_2$ sample is ~100 ps , i.e. more than one order of magnitude smaller than in 115µm-long NbN sample or 37 times smaller than in the 460 µm-long NbN sample (the same number of squares as in the $MgB_2$ sample). In order to reduce the decay time NbN nanowires to the sub-ns scale, NbN nanowires have to be very short, which requires the nanowire to be integrated with a photonic waveguide in order to provide good optical coupling. [15,16] Alternatively, several NbN nanowires could be connected in parallel, which leads to a reduced total inductance and hence to a shorter voltage decay time. [17] However, reduction of the nanowire length (aiming for a faster reset) has a negative effect on the maximum utilizable bias current for which NbN SNSPDs still do not latch into the permanent normal state ($I_{latch}/I_c$ could be as low as ~0.7).[15] Interestingly, despite a very low inductance, $MgB_2$ nanowire detectors did not show latching for bias currents up to $I_c$ . This fact can be explained by a factor of 4 times faster electron energy relaxation in $MgB_2$ (12 ps) [12] than in NbN (50ps) [18] thin films. Fast electron relaxation leads to rapid normal domain collapse and makes a fast current return from the load back into the nanowire much less critical. [19]

For an initial estimate of the $MgB_2$ photon detector response jitter, we utilized statistics for response pulse intervals collected within a 10 μs time period. Although the oscilloscope trigger was not locked to the pulsed laser, the response interval jitter is still only ~50 ps, which corresponds to a ~25 ps jitter of the rising edge (Fig.3a). This value is an upper limit for $MgB_2$ nanowire detectors, because a jitter less than that in NbN could be expected in $MgB_2$ considering a factor of 33 lower kinetic inductance.[20]

Direct demonstration of single photon detection capability requires a 100% photon coupling and a known quantum efficiency. Considering that the former is a challenging engineering task and the latter would require complex modelling, the single photon detection mode is often verified by statistical analysis of the photon counts per photon flux[4] or inter-counts intervals.[21] We chose the first method, utilizing initially a cw laser (630nm). The photon count rate $N$ vs the laser attenuation $F$ is shown in Fig. 3b at several bias currents for a 90 nm-wide and 40 μm-long nanowire. For bias currents $I_0 \geq 396$ μA, the photon count is linearly proportional to the photon flux, $N \propto F$. For smaller bias currents we observed that $N \propto F^m$ , where $m$>1, which indicates that multi-photon detection mode becomes dominant. These results show that, in contrast to NbN nanowire detectors, the single photon detection mode is limited in $MgB_2$ nanowires to a narrow bias current range in the vicinity of $I_c$ (Fig.4a). Physical modeling using accurate material parameters, is required in order to obtain a full understanding of photon detection in $MgB_2$ thin film nanowires. In this moment, we note that the aforementioned fact is probably caused by a smaller size of the normal domain (compared to the nanowire width), formed by the photon-deposited energy. Therefore, a higher dc current (vs $I_c$) is required to extend the normal domain to the edges of the nanowire. The dark count rate in the studied $MgB_2$ nanowire detectors was very small, <10 cps at the highest utilized bias current (398μA), despite the fact that the samples were not shielded from either the electrical or magnetic field interference. At this proximity to $I_c$ , the dark count rates in NbN SNSPDs often reach values of >$10^3$-$10^5$ cps. [6,22] Statistics of dark counts could be utilized to analyze detection principles, [22,23] however in the $MgB_2$ nanowire case it would presume dark count integration over much longer periods.

Similar results were obtained for detection of IR photons ($\lambda$=1550 nm) from the 80 fs pulsed laser by a 35 nm × 120 μm $MgB_2$ sample ($I_c$=97 μA). For $I_0$ >0.97×$I_c$, a linear $N(F)$ dependence (Fig.4b) indicates that the detector response is triggered by single photon absorption event. For $I_0$ <0.97×$I_c$, a combination of both 1- and 2- photon responses was observed ($N \propto F^m$, $m$>1). At a high photon flux ($F$>0.1), i.e. at a large number of photons per pulse, multi-photon response even dominates at the highest bias current, as has been observed in NbN SNSPDs [24] for large laser power.

In conclusion, in a comparative experiment we demonstrated that reset time in 120 μm-long $MgB_2$ nanowire photon detectors is ~100 ps, i.e. much shorter than that of their NbN counterparts.

Despite a low inductance, $MgB_2$ photo detectors self-reset (i.e. they do not latch) at bias currents up to $I_c$ . Our results show that $MgB_2$ nanowire photon detectors are capable of single photon detection for both visible and communication (1550 nm) spectral ranges, combined with a low dark count rate, hence forming a technological platform for sensitive and high speed photon detectors with a large cross-section and a low jitter.

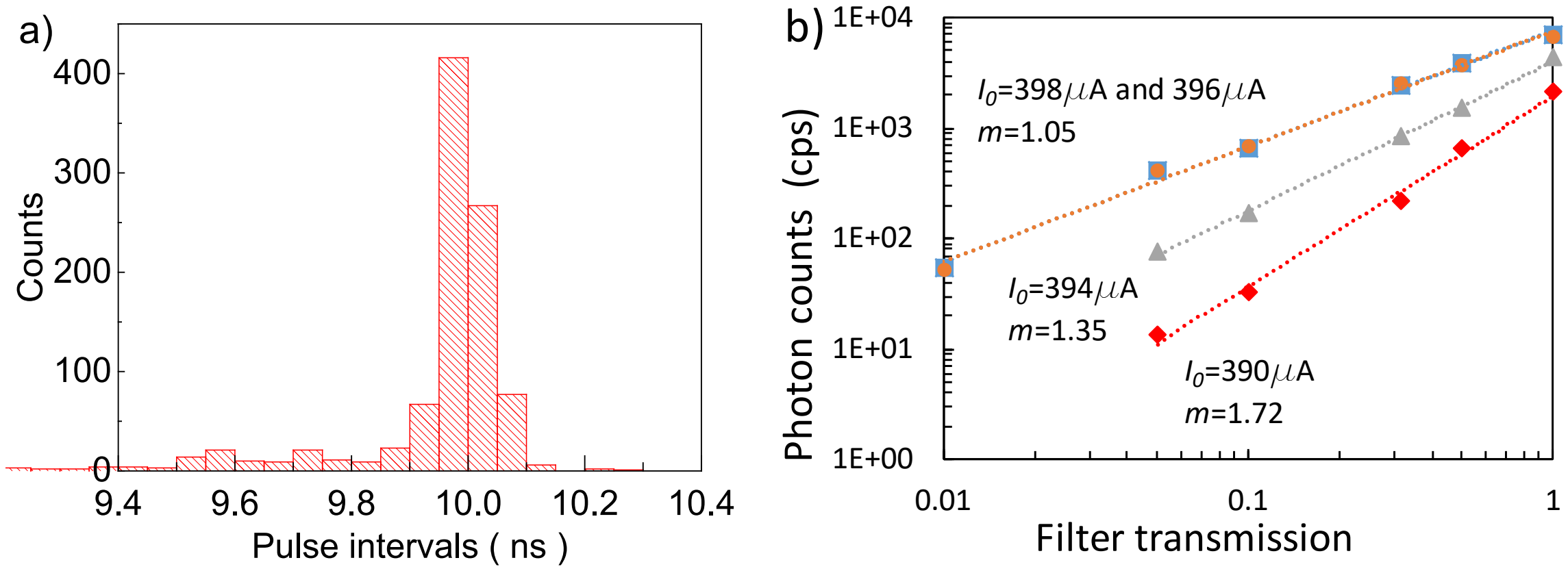

Fig. 3. (a) Statistics for the response intervals (time span of 10 μs) for the $MgB_2$ nanowire as in Fig.2b. The laser repetition rate was 100 MHz. (b) Photon counts vs relative photon flux (filter transmission with different Optical Densities OD). The $MgB_2$ detector (90 nm × 40 μm, $I_c$=399μA) was illuminated with a cw laser (630 nm). Lines are $N \propto F^m$ fits. All measurements were done at 5K.

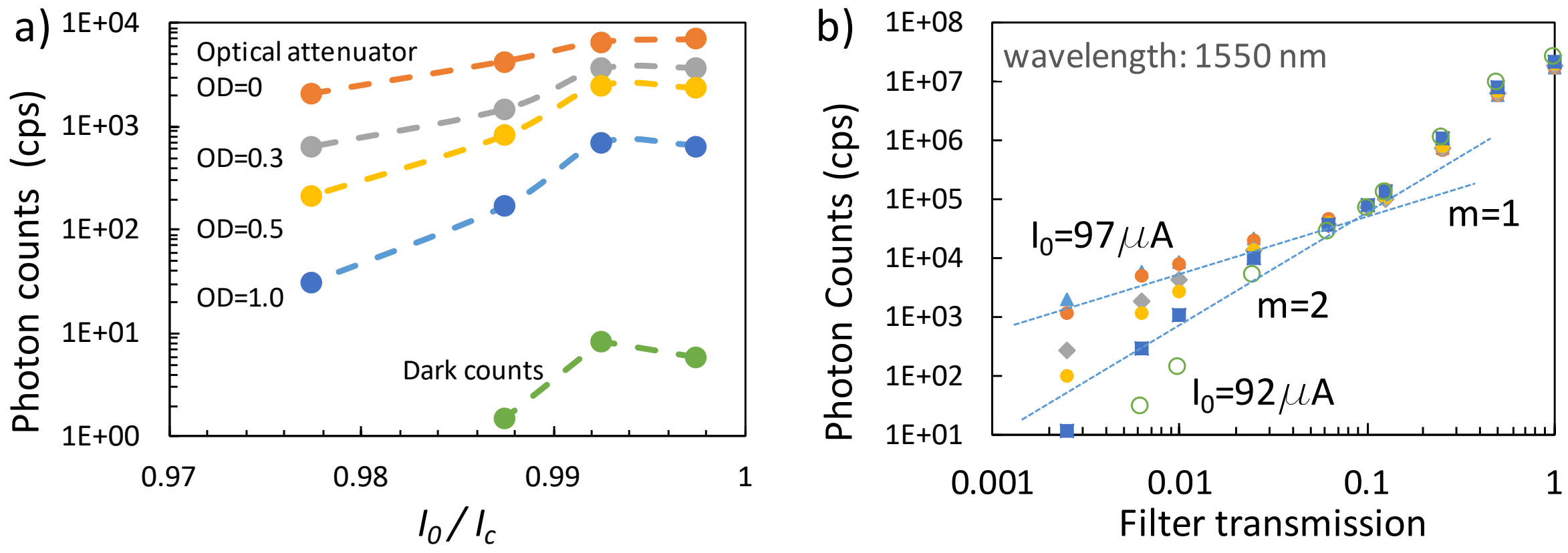

Fig.4 (a) Photon (630nm) counts vs normalized bias current, for the same sample and the laser as in (Fig.3b); b) Photon counts vs photon flux for a 35nm-wide and 120μm-long nanowire ($I_c$=97μA) illuminated with a pulsed laser (80 fs , 1550 nm, 100 MHz repetition rate). Lines are $N \propto F^m$ fits. All measurements were made at 5K.

ACKNOWLEDGMENT

$MgB_2$ film technology was developed under European Research Council grant #308130-Teramix. This research was also supported by Swedish Research Council (grant # 2016-04198) and by the Swedish National Space Agency (grant # 198/16). The work was conducted using facilities at the Nanofabrication Laboratory and the Kollberg Laboratory of Chalmers University of Technology.

Supplementary Materials.

# IR and visible range single photon detection in superconducting $MgB_2$ nanowires.

*Sergey Cherednichenko*, Narendra Acharya, Evgenii Novoselov, and Vladimir Drakinskiy*

Terahertz and Millimetre Wave Laboratory, Department of Microtechnology and Nanoscience, Chalmers University of Technology, SE-412 96, Gothenburg, Sweden

* *Corresponding author serguei@chalmers.se*

## NbN film deposition, NbN nanowire fabrication NbN kinetic inductance measurements.

NbN films were deposited by DC reactive magnetron sputtering in an Ar + $N_2$ environment on 330µm thick C-plane double side polished sapphire substrates maintained at 800 °C. The film was patterned into 100nm wide (100nm spacing) meandering nanowires with Reactive Ion Etching ($CF_4$:$O_2$, 5:1 flow ratio) through a mask defined by e-beam lithography on a positive (PMMA A2, 60nm thick) resist. During the RIE process, the total pressure was kept as low as possible (3.8mTorr being the lowest used to provide a stable plasma discharge) in order to increase the etching anisotropy. Electrical contact pads (as single-port 50Ω coplanar waveguide, CPW) were fabricated in 10nm Ti / 200nm Au / 40 nm Ti using a standard lift-off process. The NbN film between the CPW contacts was removed by means of wet-etching (nitric acid/ hydrofluoric acid/ acetic acid, 5:3:3) with a resistive mask placed over the nanowires (photo resist AZ1512).

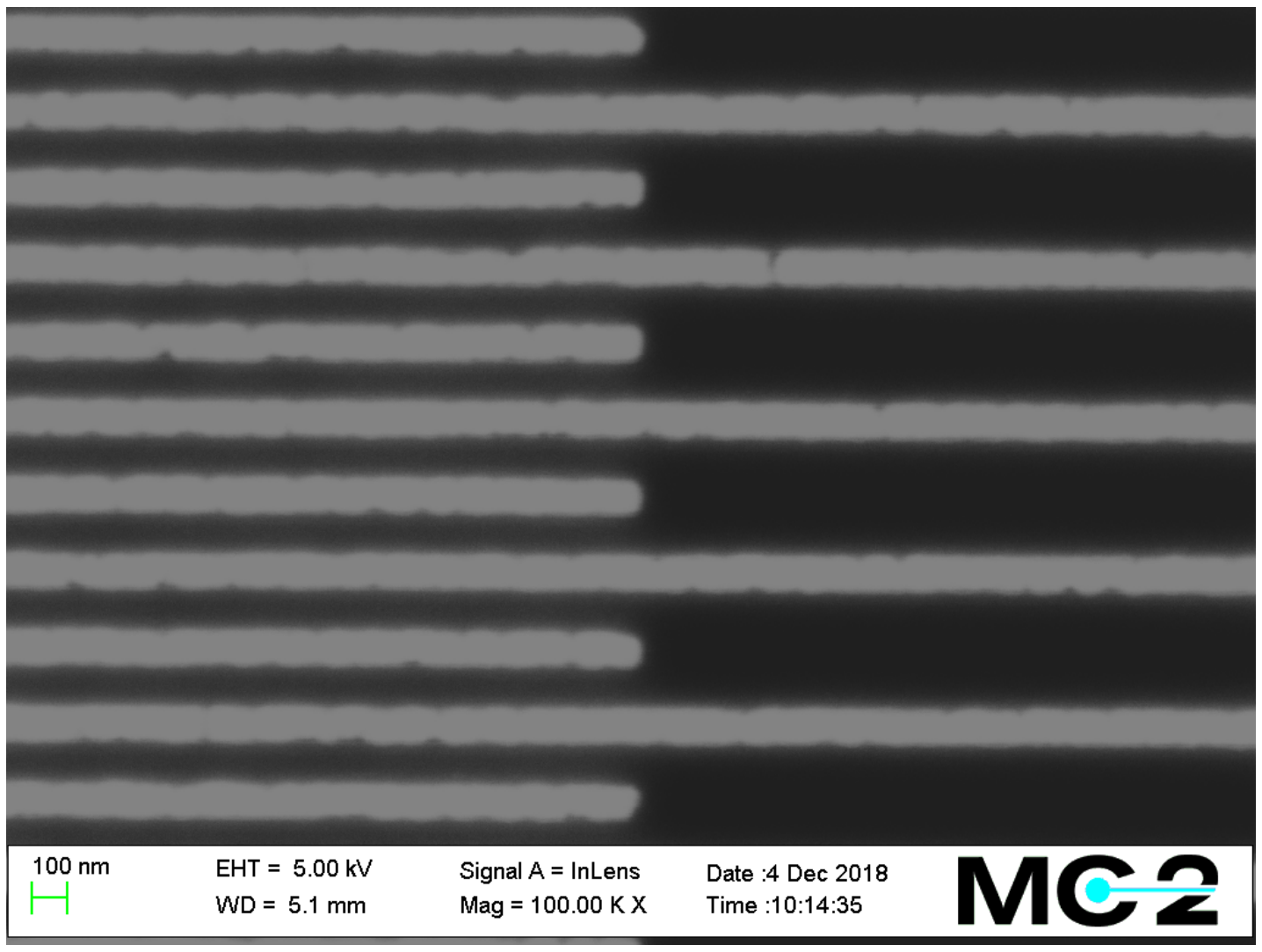

FIG. S1 SEM image of a section of a NbN (dark area) nanowire.

## NbN kinetic inductance measurements.

Kinetic inductance of $MgB_2$ nanowires in the superconducting state (4.8 K -32 K) was obtained by fitting the measured microwave impedance to $Z(f)=2\pi \times f \times L$, where $f$ is the probing microwave frequency and $L$ is the sample inductance. The microwave impedance was measured using a Vector Network Analyzer (VNA) (Rohde&Schwarz ZVA 67, 10MHz-67GHz) in the single port configuration. The NbN device wafer was cooled with a cryogen-free probe station (Lakeshore CRX-4K). NbN nanowire samples, integrated with coplanar waveguide contacts made of Ti/Au film, were contacted via a high frequency GSG probe on a XYZ manipulator stage. Other measurement details were the same as in Ref. S1.

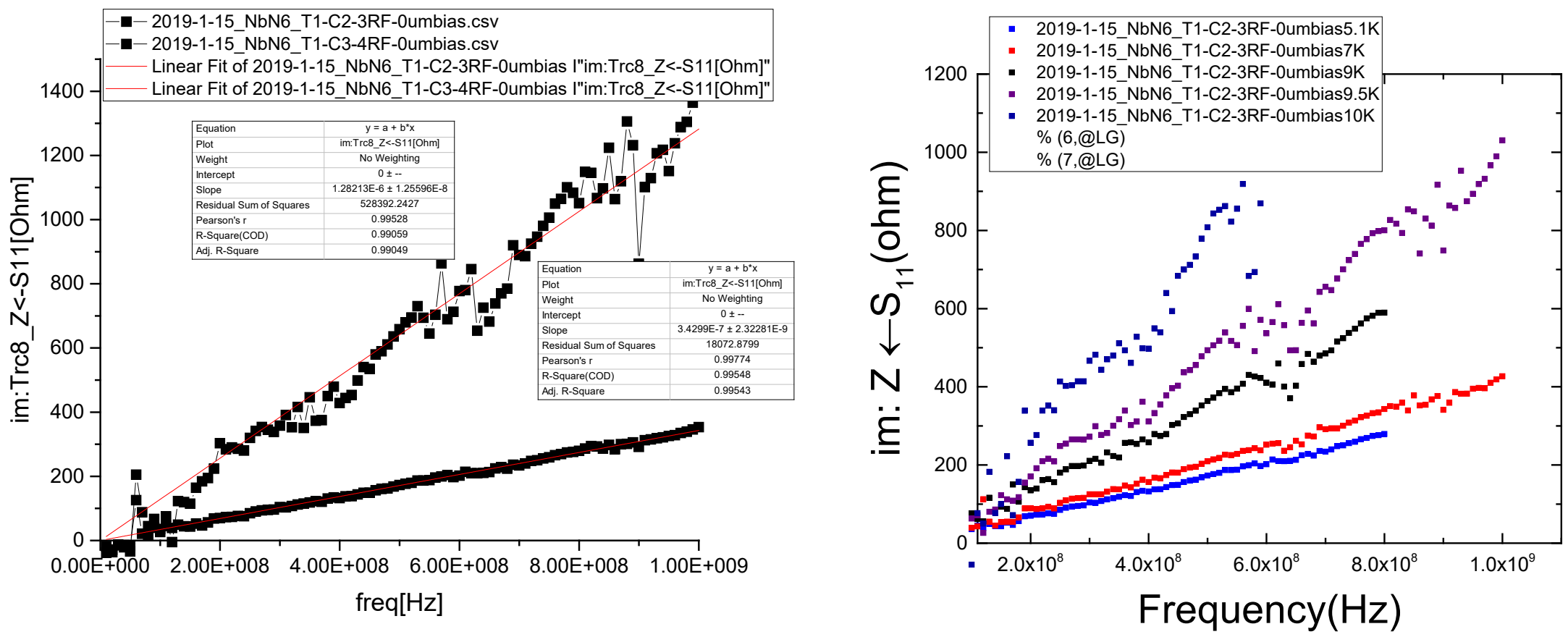

FIG. S2 Imaginary impedance as measured at 5K at zero bias for NbN meanders. For different meander lengths (left), and for a single meander with varying temperature (right). Linear fits to the impedance were used in order to extract kinetic inductance. The total kinetic inductance for sample T1-C2 (w×L=100nm×460 μm) is 200nH, and for sampleT1-C3 (w×L=100nm×115 μm) is 54nH, as it is calculated from the expression for the inductive impedance ($Z=2\pi\times f\times L_k$) ($f$ is the linear frequency).

## Optical response measurements

The optical response measurements of the devices were taken in the same cryogenic probe station as for the kinetic inductance study. Lasers illuminated the samples through a quartz pressure window. The standard IR filter (not transparent for 1550nm photons) on the 60K shield was replaced with a WG12012-C window (Thorlabs). The output beam from the cw λ=630nm laser diode module required no extra collimation, resulting in a light spot diameter of 3-4mm. The IR (λ=1550nm) 80-fs pulse laser (Toptica) is fiber-coupled and a fiber launcher (FIG.S3) was utilized to collimate the IR laser beam. The IR-laser spot size (on the sample) was approximately 5mm, verified using SNSPD photo counts vs the laser XY-shift. Two sets of light attenuators were utilized for the visible and IR light. A bias-T (as described in the main text) and low noise amplifiers were placed in room temperature.

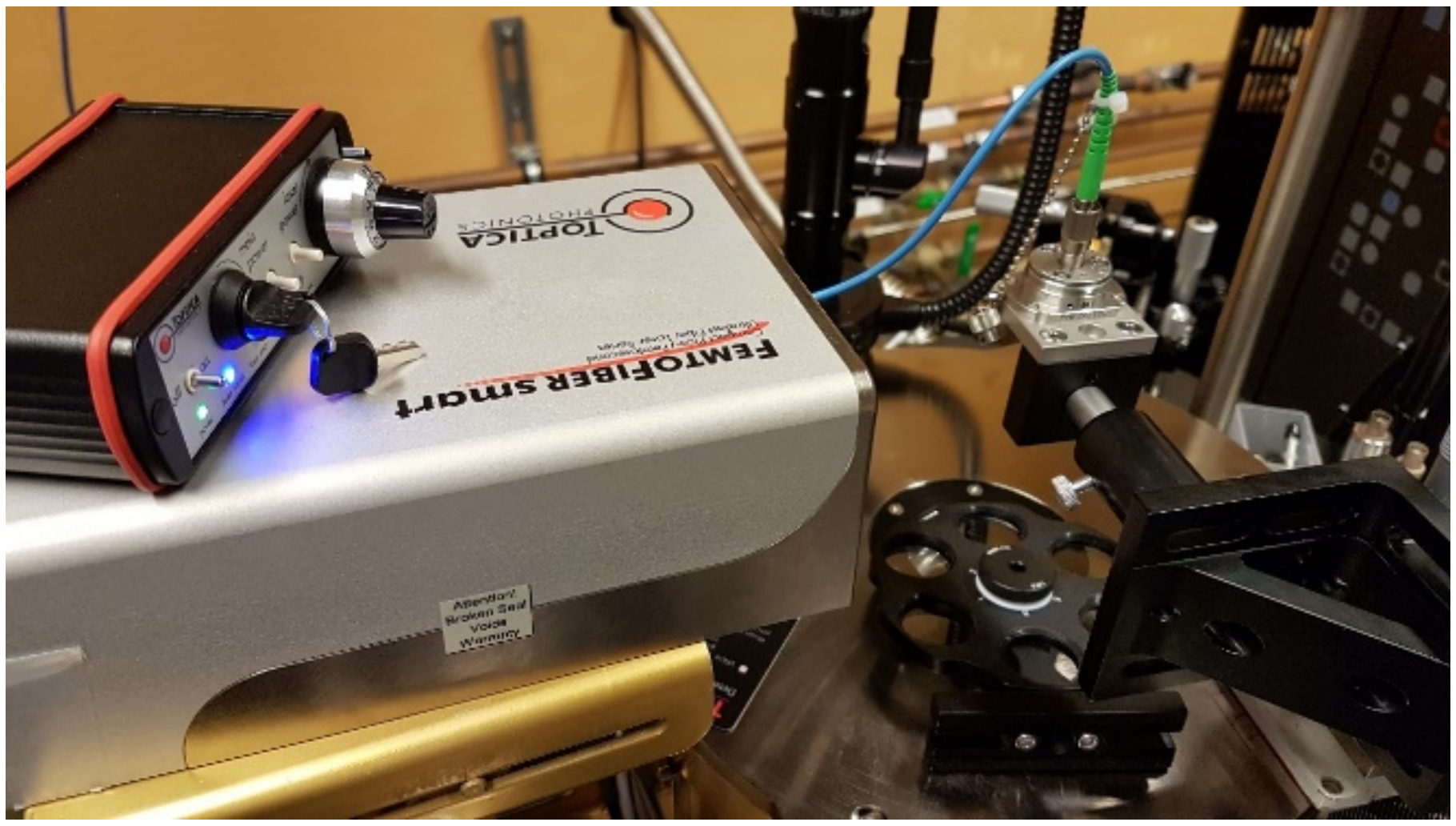

FIG. S3 The optical characterization set-up. The fs-laser, the fiber collimator, and the set of attenuators (in a rotation holder) positioned over the window in the cryogenic probe-station.